\documentclass[
	twocolumn,
	amsmath,
	amssymb,
	prl,
	aps,
 superscriptaddress
	]{revtex4-2}
\usepackage{graphicx}
\usepackage{xcolor}
\usepackage{amsmath}
\usepackage{amssymb}
\usepackage{booktabs}
\usepackage{titlesec}
\usepackage{multirow}
\usepackage{lipsum}
\usepackage{comment}
\usepackage{soulutf8}
\usepackage{ulem}
\usepackage{color}

\sethlcolor{}

\begin{document}

\title{Spatiotemporal Hierarchy of Slow Avalanches During Creep}

\author{Vladimir Yu. Rudyak}
\affiliation{Department of Condensed Matter, School of Physics and Astronomy, Tel Aviv University, Tel Aviv 69978, Israel}
\author{Dor Shohat}
\affiliation{Department of Condensed Matter, School of Physics and Astronomy, Tel Aviv University, Tel Aviv 69978, Israel}
\author{Yoav Lahini}
\affiliation{Department of Condensed Matter, School of Physics and Astronomy, Tel Aviv University, Tel Aviv 69978, Israel}

\begin{abstract}
Far from equilibrium, amorphous solids exhibit structural relaxations that span a vast range of timescales such as physical aging and creep. Recently, it has been shown that such relaxations are driven by via intermittent, scale-free, yet anomalously slow cascades of local rearrangements, termed ‘thermal avalanches.’ Here, we investigate the spatio-temporal dynamics of these avalanches during logarithmic creep, using simulations of a model amorphous solid. By systematically disentangling mechanical and thermal activation events, we reveal that thermal avalanches have a hierarchical spatio-temporal structure: localized rearrangement events group into fast and compact cascades, which then promote the thermal activation of subsequent cascades via long-range, noise-mediated facilitation. This process results in heavy-tailed temporal correlations reminiscent of seismic activity. We validate these findings using experiments on slow relaxation of crumpled matter. Our work provides a framework for identifying noise-mediated correlations and elucidates the rich structural dynamics underlying slow relaxation of amorphous solids.

\end{abstract}

\maketitle

The dynamics of disordered systems is dominated by intermittent bursts of activity. This hallmark is observed across a wide range of systems, from spin glasses \cite{perkovic1995avalanches,pazmandi1999self} and deformed metals \cite{antonaglia2014bulk}, through granular and porous materials \cite{shang2020elastic,denisov2016universality,baro2013statistical}, to neuronal activity in the brain \cite{beggs2003neuronal,friedman2012universal}. 
Due to the tendency of disordered systems to lie on the verge of instability \cite{muller2015marginal,lin2016mean}, an abrupt event – be it spin flipping, neuron firing, or localized plastic deformation – can trigger other events in a rapid sequence, forming an avalanche. These avalanches are typically scale-free and system-spanning \cite{bak1987self}, and are therefore crucial for dynamics yet difficult to predict.

Strikingly, avalanche dynamics is strongly altered in the presence of a minute temperature or noise. Here, a new triggering mechanism arises, based on facilitation and thermal activation \cite{Ozawa2023elasticity,tahaei2023scaling,korchinski2024microscopic,ferrero2017spatiotemporal,guiselin2021microscopic}. An instability can either trigger the next event immediately; or lower its energy barrier, thus facilitating its stochastic noise-driven activation at a later time. Highly correlated yet anomalously slow sequences of such events, termed 'thermal avalanches', have recently been shown to play a crucial role in the heterogeneous relaxation of glasses \cite{tahaei2023scaling,gavazzoni2023testing}, and in the creep of amorphous solids \cite{shohat2023logarithmic, korchinski2024microscopic} and disordered interfaces \cite{durin2023,de2024dynamical}. Despite these advances, a complete description of the spatio-temporal structure of thermal avalanches is missing. Resolving this structure may help clarify how relaxation, yielding, and flow develop in amorphous matter across scales \cite{nicolas2018deformation}.

In this work, we uncover the rich spatiotemporal dynamics of thermal avalanches using simulations of a model amorphous solid -- a disordered network of bi-stable elastic bonds \cite{shohat2025emergent}. Under an external load and at low temperature, the network self-organizes into a marginally stable state \cite{shohat2025emergent} and exhibits logarithmic creep driven by thermal avalanches \cite{shohat2023logarithmic}. By analyzing the spatio-temporal correlations between events, we systematically disentangle mechanical and thermal activation events and reveal a complex, hierarchical structure: a localized instability triggers others mechanically, forming fast and compact cascades; each cascade facilitates the delayed thermal activation of other cascdes, giving rise to heavy-tailed Omori-like temporal correlations \cite{senshu1959time,bak2002unified,korchinski2024microscopic} and long-ranged spatial correlations, forming a thermal avalanche; consecutive avalanches exhibit anomalous temporal correlations, akin to those observed between earthquakes \cite{corral2003local,talbi2010mixed,li2024double}. Finally, we show evidence for this hierarchy in experiments with logarithmic creep of crumpled thin sheets under load \cite{shohat2023logarithmic}. Our study sheds light on the rich dynamics that underlies slow relaxations in amorphous solids \cite{korchinski2024microscopic,popovic2022scaling,bouttes2013creep,cottrell1952time}. It also establishes a framework to identify facilitation-driven correlations and detect thermal avalanches in potentially noisy, imprecise, or coarse-grained data, in systems ranging from glass \cite{bhaumik2022avalanches,takaha2024avalanche} to seismic faults \cite{zaliapin2016global,houdoux2021micro}.

\section{Disordered network model}
\begin{figure}[tbp]
 \centering
 \includegraphics[width=0.85\linewidth]{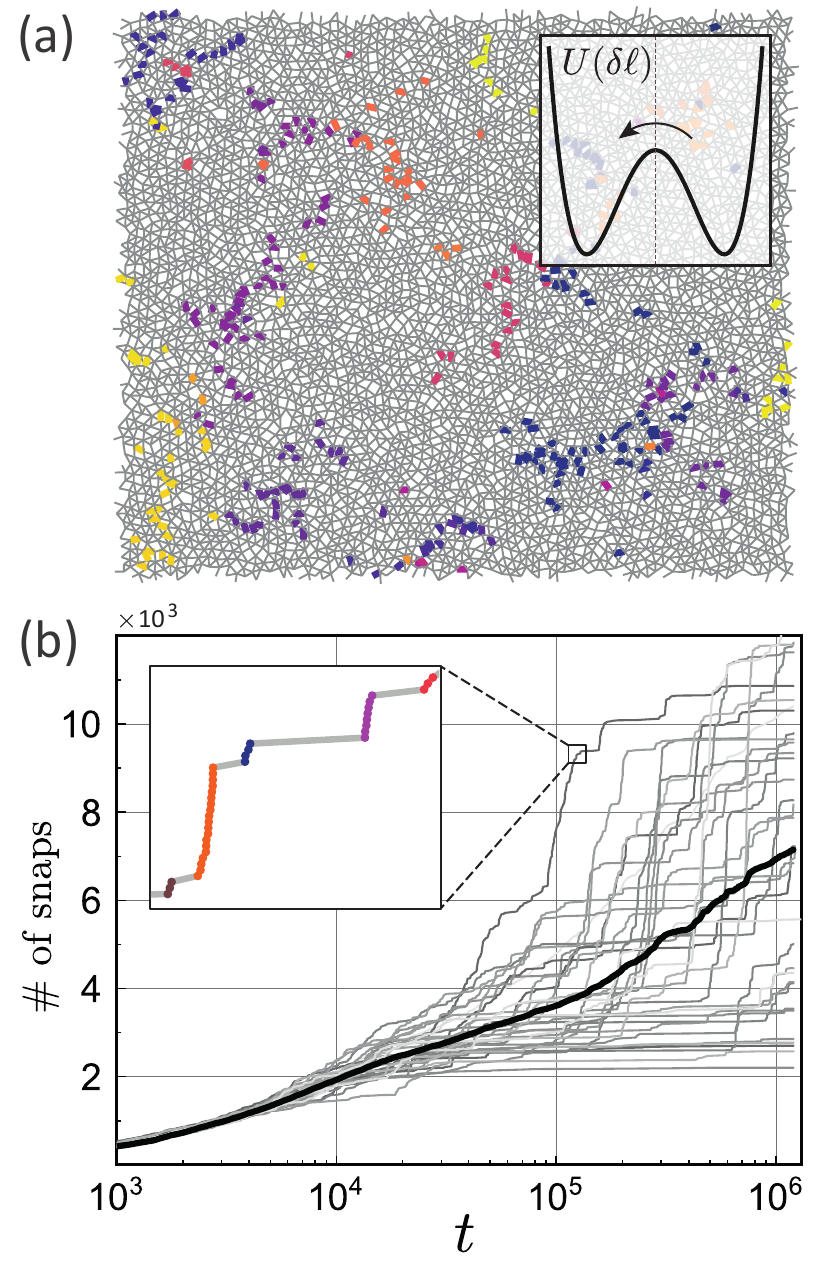}
 \caption{\textbf{Avalanches during creep -- }
 (a) A disordered network of bi-stable bonds (partial view). A single, system spanning thermal avalanche is highlighted, with different colors representing cascades of snaps; Inset: the double-well potential of each bond, as a function of its displacement $\delta\ell=\ell-\ell^{(0)}$; (b) Cumulative number of bond snaps over time for 40 realizations (gray), and their average (black line). The evolution is logarithmic and driven by abrupt avalanches. Inset: Zooming in on a single avalanche reveals distinct cascades.
 }
 \label{fig:system}
\end{figure}

We consider a minimal structural model for amorphous solids, a disordered network of $N_b\approx10^5$ bi-stable elastic bonds (Fig.\,\ref{fig:system}a) \cite{shohat2022memory, shohat2023logarithmic,yan2013glass}. Each bond $i$ in the network is characterized by a length $\ell_i$ and a double-well potential of the form $U_i(\ell_i)=\frac{a_4}{4}(\ell_i-\ell^{(0)}_i)^4 -\frac{a_2}{2}(\ell_i-\ell^{(0)}_i)^2$, where the rest length $\ell^{(0)}_i$ is randomized, and $a_2,a_4$ are network constants (see Appendix for details). In the following, we normalize all distances by the mean rest length $\langle \ell^{(0)}\rangle$. An important aspect of this model is that due to the incompatibility between the rest lengths of neighboring bonds, they carry excess stress, which can significantly lower their effective activation energy barriers \cite{shohat2025emergent}. 

This model was shown to capture several hallmark behaviors of amorphous solids, from memory effects under cyclic driving \cite{shohat2022memory}, creep and physical aging \cite{shohat2023logarithmic,shohat2025aging}, to marginal stability \cite{shohat2025emergent}. In particular, under constant external load and at low temperatures, the network exhibits a slow logarithmic compaction driven by discrete instabilities: abrupt events in which a bond snaps between the two wells of its potential. These instabilities trigger and facilitate each other, forming scale-free, slow thermal avalanches \cite{shohat2023logarithmic,korchinski2024microscopic}.

Here, we simulate the logarithmic creep dynamics of such networks using LAMMPS \cite{thompson2022lammps}. The Langevin simulations consider a very low temperature $T$ and a constant external force $F$ applied at the boundaries, resulting in slow, intermittent compaction via thermal avalanches \cite{shohat2023logarithmic} (Fig.\,\ref{fig:system}b). During the simulation, our custom code tracks the state of each bond, and the full network configuration is registered when and only when an instability occurs (see Appendix and \cite{lammps-download}). This enables tracking individual bond snap events with single-time-step accuracy over extremely long simulation times, increasing the temporal resolution compared to previous studies \cite{shohat2023logarithmic}. Below, we leverage this ability to shed light on the complex spatiotemporal dynamics of thermal avalanches.

\begin{figure}[tbp]
 \centering
 \includegraphics[width=0.99\linewidth]{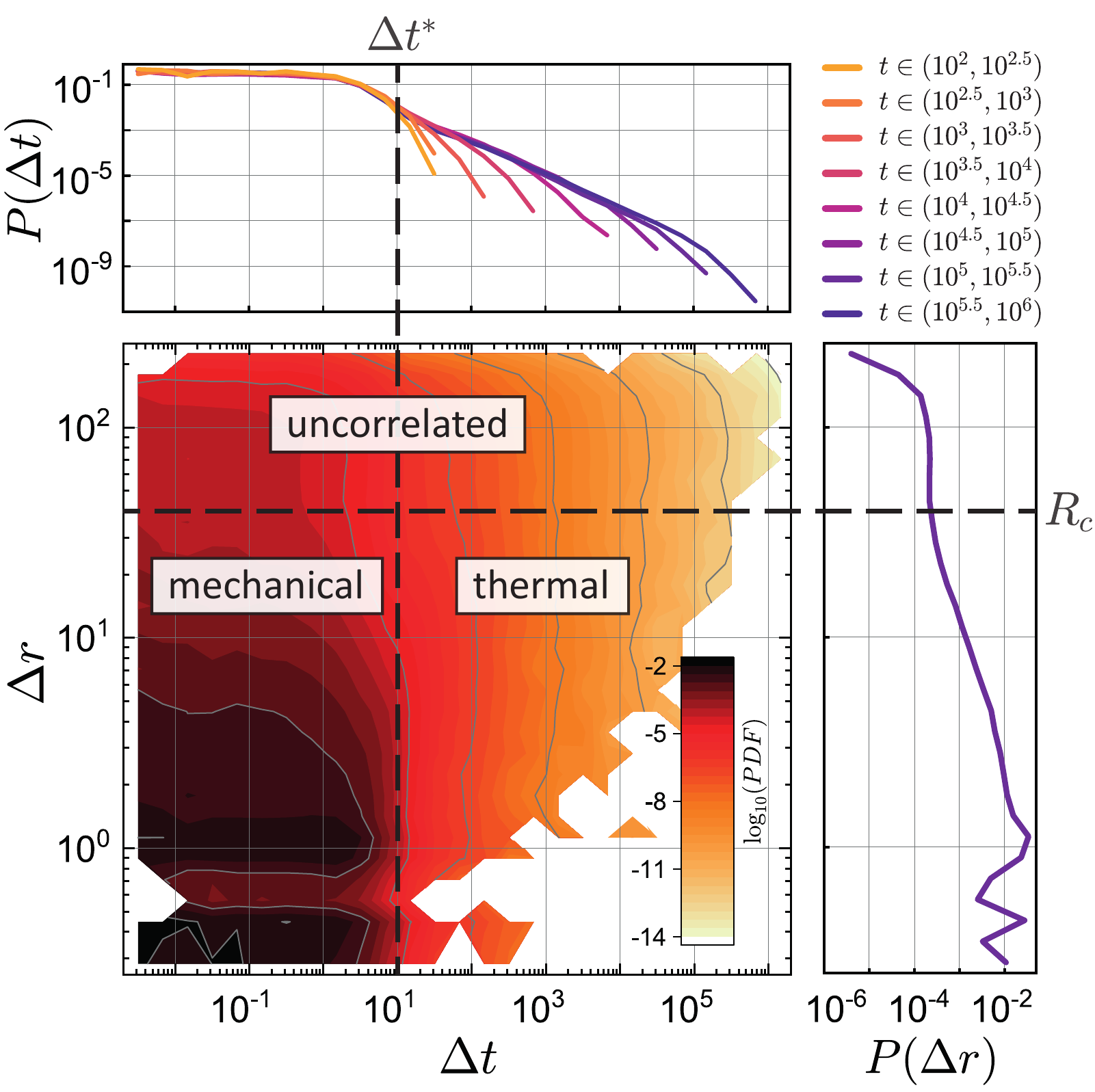}
 \caption{\textbf{Spatiotemporal correlations -- }
 Joint probability distribution $P(\Delta t,\Delta r)$ of the waiting times and distances between consecutive events, divided to three regimes (main panel). The overall distribution of waiting times (top panel) distinguishes between mechanical triggering at short times, and age-dependent thermal triggering at long times. The overall distribution of distances identifies the spatial correlation length $R_c$.
 }
 \label{fig:bondsnaps}
\end{figure}

\section{Elementary instabilities}
We begin by considering the distribution of waiting times between subsequent bond snaps $P(\Delta t_i=t_i-t_{i-1})$. A set of these distributions, calculated at different ages of the system (defined as the time elapsed since the beginning of the simulation) is presented in the top panel of Fig.\,\ref{fig:bondsnaps}. These distributions reveal two distinct behaviors. At short waiting times $\Delta t_i<10$, all distributions are flat, and collapse on top of each other, meaning that the behavior at short waiting times is age independent. In contrast, for $\Delta t>10$, distributions taken at different ages deviate from each other. In particular, longer waiting times are observed, meaning that dynamics slows with the system's age. As we show below, these distinct behaviors are associated with different triggering mechanisms. 

We next consider the spatial correlations between consecutive instabilties, $P(\Delta r)$, where $\Delta r_i=|\mathbf{r}_i-\mathbf{r}_{i-1}|$ is the distance between the snapping bonds, shown on the right panel in Fig.\,\ref{fig:bondsnaps}. Here as well it is possible to distinguish between decaying spatial correlations for short to intermediate distances, and a flat distrbution (decorrelation) at long distances. 

We combine the results and examine the spatio-temporal distribution of subsequent snaps $P(\Delta t_i,\Delta r_i)$, presented in the main panel of Fig.\,\ref{fig:bondsnaps}. In the following, we will use these data to identify three dynamical regimes: direct mechanical triggering, correlated thermal activations, and uncorrelated events. 

The first regime occurs at short inter-event waiting times $\Delta t_i<\Delta t^*=10$, and short to intermediate inter-event distances $\Delta r_i<R_\text{c}$ ($R_\text{c}\approx40$ bond lengths). This regime represents an almost immediate destabilization of a bond due to the snapping of another bond nearby \cite{shohat2025emergent}, corresponding to direct mechanical triggering. As discussed above, the waiting time distribution in this regime is age-independent -- dynamics of mechanical triggering is age-independent. 

The second regime involves longer waiting times $\Delta t_i>\Delta t^*$. These have been shown to arise from facilitated thermal activations. Here, the stress redistribution induced by a snapping bond lowers the effective energy barrier of other bi-stable bonds, thereby facilitating their thermal activation at a later time \cite{Ozawa2023elasticity, shohat2023logarithmic, korchinski2024microscopic}. As discussed above and shown in Fig.\,\ref{fig:bondsnaps}, in this regime the distribution of waiting times between events follows a power-law, and is age-dependent: as the system ages, events become correlated across longer and longer times. We stress that the transition between the two regimes is not chosen arbitrarily. In Appendix \ref{sec:appB}, we provide additional measurements which confirm that $\Delta t^*$ separates two distinct triggering processes. Finally, above the correlation length $R_\text{c}$, $P(\Delta r_i)$ reaches the background noise plateau, indicating uncorrelated activity. 

Having identified these three regimes, we will now construct a bottom-up understanding of thermal avalanches.


\section{Athermal cascades}
\begin{figure}[tbp]
 \centering
 \includegraphics[width=1\linewidth]{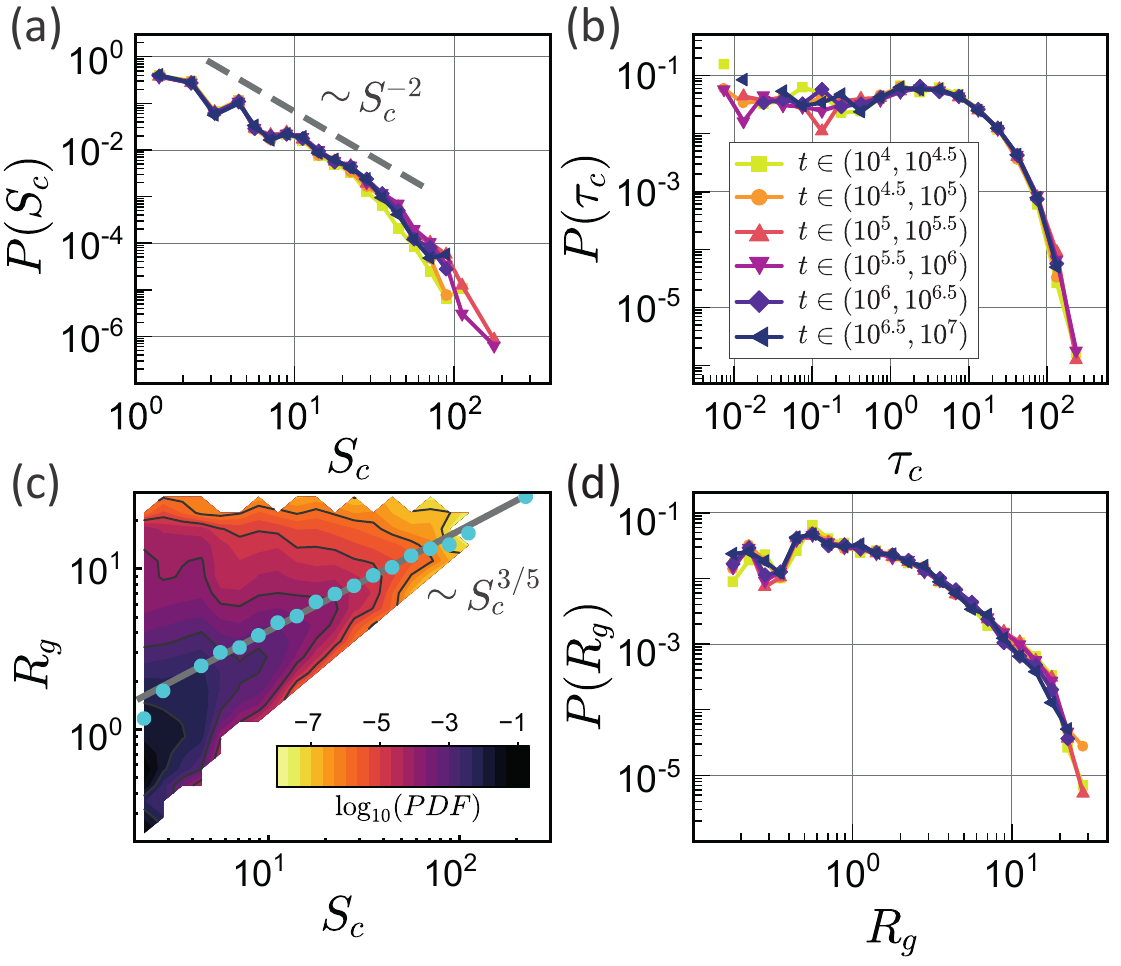}
 \caption{\textbf{Mechanical cascades -- }
 (a) Distribution of cascade magnitudes $S_c$ exhibiting a power law; (b) Distribution of cascade durations $\tau_c$, which is age-independent; (c) Distribution of radii of gyration $R_g$ as a function of $S_c$, and their mean (teal points); (d) $R_g$ probability distribution.
 }
 \label{fig:cascades}
\end{figure}

We start by characterizing the fast sequences of mechanically triggered snaps, which we term cascades. To track these cadcades, we scan all events one by one, and add them to the spatially closest cascade that contains a preceding event with $\Delta t_i<\Delta t^*$ and $\Delta r_i<R_c$. Otherwise, the snapping event is considered the first in a new cascade. This process groups snapping events into cascades, even if those coexist in different regions of the network. 
An example of this grouping process is shown in Fig.\,\ref{fig:system}a, where different cascades are marked with different colors. The magnitude of each cascade $S_c$, defined as the number of snapping events it contains, is approximately power-law distributed $P(S_c)\sim S_c^{-\alpha}$ with $\alpha\approx2$, as shown in Fig.\,\ref{fig:cascades}a.

Next, we compute the duration of each cascade $\tau_c$, defined as the time elapsed between the first and last events. We find that the duration distribution $P(\tau_c)$ is flat, compact and age-independent (Fig.\,\ref{fig:cascades}b).

Spatially, we observe that cascades exhibit a range of morphologies; some are line-like while others resemble compact blobs (see SI). To characterize these spatial structures, we compute the radius of gyration $R_g^2=\frac{1}{S_c}\sum_{i=1}^{S_c}(\mathbf{r}_i-\mathbf{r}_\text{CM})^2$, where $\mathbf{r}_\text{CM}$ is the center of mass of the cascade. On average, we find a scaling of $R_g\sim S_c^{3/5}$ (Fig.\,\ref{fig:cascades}c-d). This indicates that cascades occupy more space than a standard random walk, possibly due to self-avoidance effects. Together, these measurements are consistent with the notion that cascades are driven by local mechanical destabilization, in an age-independent manner.

\section{Thermal avalanches}
We now consider groups of events which are correlated across long time-scales. Here, an event does not trigger others immediately. Rather, due to long-range elastic interactions \cite{shohat2025emergent}, it can lower the energy barriers of other events, thereby facilitating their thermal activation, which can occur at a later time \cite{shohat2023logarithmic, korchinski2024microscopic}. This process results in correlated activity that spans many time scales in a manner that is age-dependent due to the growth of effective energy barriers, as shown for $\Delta t > \Delta t^*$ in the top panel of Fig. \ref{fig:bondsnaps}

\begin{figure}[tbp]
 \centering
 \includegraphics[width=1\linewidth]{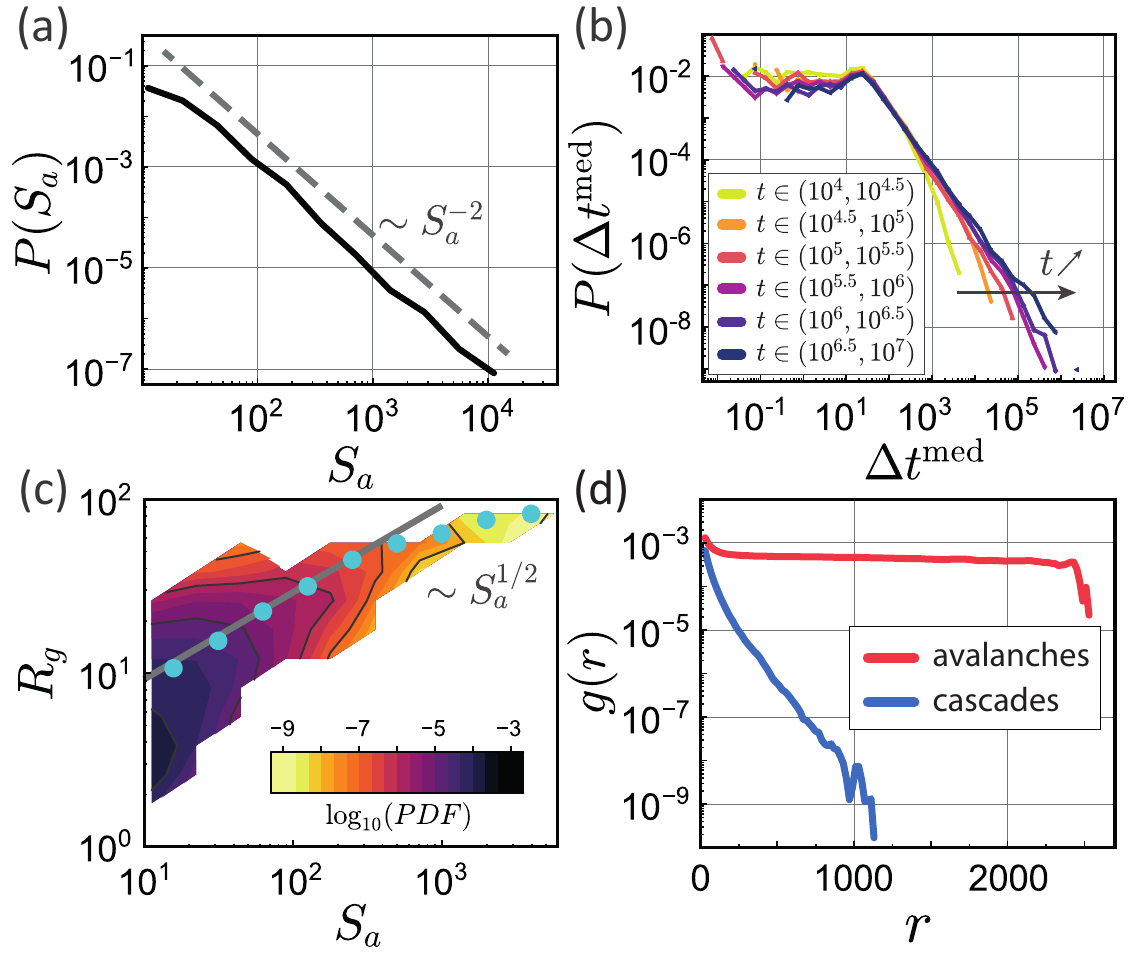}
 \caption{\textbf{Thermal avalanches -- }
 (a) Distribution of avalanche magnitudes $S_a$ exhibiting a power law; (b) Distribution of waiting times between cascades  $\Delta t^\text{med}$, with a power-law tail that extends and exhibits aging; (c) Distribution of radii of gyration $R_g$ as a function of $S_a$, and their mean (teal points); (d) Radial distribution function $g(r)$ for both cascades and avalanches. Cascades are localized while avalanches span the whole network. 
 }
 \label{fig:intercascade}
\end{figure}

As the short time correlations correspond to the mechanical cascades described above, we can simplify the analysis by considering spatiotemporal correlations between cascades. To this end, we coarsen cascades to point events in time (using their median time $t^\text{med}$) and space (center of mass $\mathbf{r}_\text{CM}$). Indeed, the distribution of waiting times between the coarsened cascades $P(\Delta t^\text{med})$, shown in Fig.\,\ref{fig:intercascade}b, captures the age-dependent, power-law tail observed at the single snap level in Fig.\,\ref{fig:bondsnaps}.

To group cascades into avalanches and determine the spatiotemporal correlations between them, we use an algorithm similar to the one described above for single snaps. However, since the network exhibits aging, and the typical waiting time between correlated cascades grows linearly in time \cite{shohat2023logarithmic, korchinski2024microscopic}, we must consider a varying timescale. To account for this slowdown, we operate in log-time and consider two subsequent cascades as part of the same avalanche if $\Delta(\log t^\text{med})=\log t_{i}^\text{med}-\log t_{i-1}^\text{med}<\Theta$ and $\Delta r_\text{CM}<R_a$, where $\Theta$ is a constant above which the correlation drops sharply (see SI). Here we use the same spatial correlation length $R_a\sim40$ bond lengths \footnote{Alternatively, choosing a correlation length that grows with the speed of sound in the system yields similar results, as we discuss in the SI.}. Additionally, to reduce noise, we ignore cascades that consist of a single snap.

The resulting avalanche magnitudes $S_a$, representing the total number of snaps, are power law distributed with a similar exponent to the cascade magnitudes $P(S_a)\sim S_a^{-2}$ (Fig.\,\ref{fig:intercascade}a). Yet, in contrast to cascades, thermal avalanches are not compact; an avalanche of magnitude $S$ covers a larger area than that of a single cascade of the same magnitude (see radii of gyration $R_g$ in Figs.\,\ref{fig:cascades}c and \ref{fig:intercascade}c). $R_g$ scales with the avalanche magnitude as $R_g\sim S_a^{1/2}$, before reaching system size. This implies a fractal dimension of approximately $d_f\approx2$ for large avalanches, consistent with previous work \cite{tahaei2023scaling} (see SI). To demonstrate that avalanches are system-spanning, we also show the radial distribution function $g(r)=\frac{1}{2\pi r \rho}\langle\sum \delta(\boldsymbol{r}-\boldsymbol{r}_i)\rangle$, where $\rho$ is the mean number of snaps per area (Fig.\,\ref{fig:intercascade}d).


\section{Temporal hierarchy}
\begin{figure}[tbp]
 \centering
 \includegraphics[width=1\linewidth]{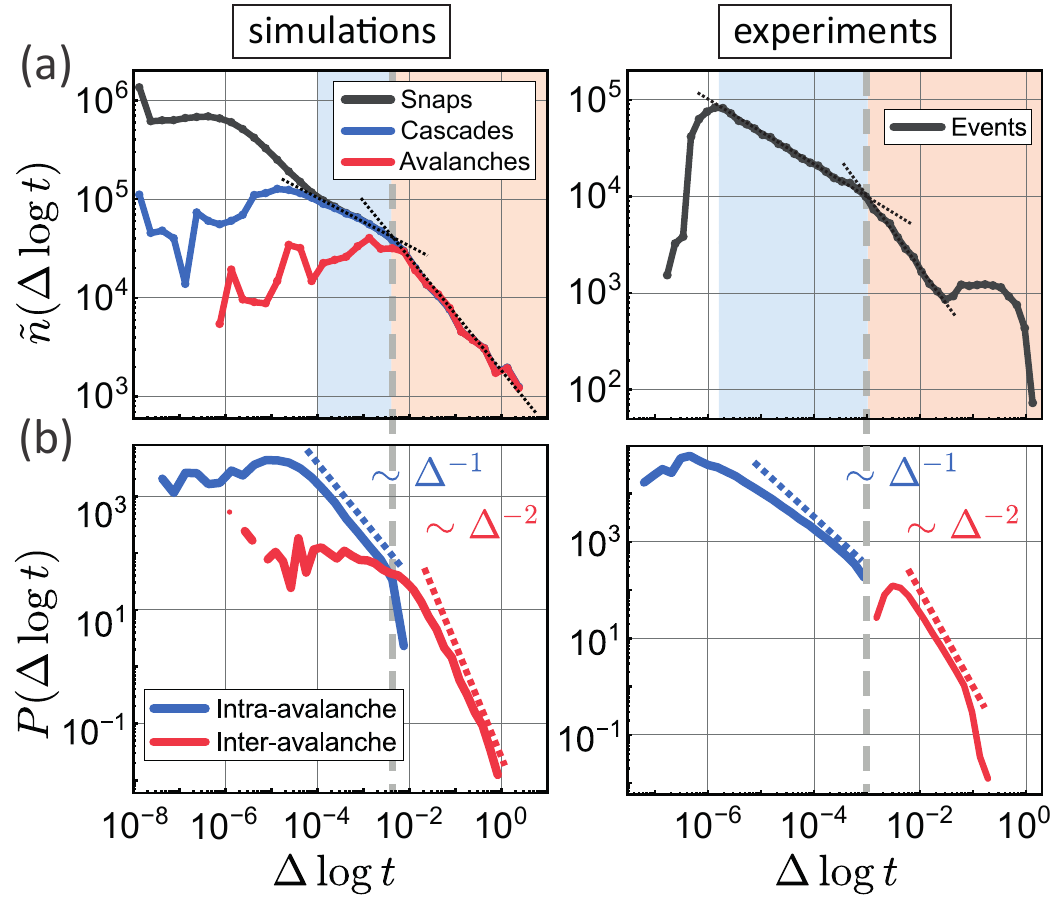}
 \caption{\textbf{Temporal hierarchy -- }
(a) Conditional event rate $\tilde{n}$ as a function of the log-waiting time, with two regimes representing cascades and avalanches in simulations (left). The same regimes are identified in the experimental data of Ref. \cite{shohat2023logarithmic} (right); (b) Probability distribution of log-waiting times, where power laws $\sim \Delta^{-1}$ and $\sim \Delta^{-2}$ characterize intra- and inter-avalanche activity, respectively.
 }
 \label{fig:avalanches}
\end{figure}

The temporal hierarchy of mechanical and thermal correlations is most clear when one analyzes the conditional snap density in log-time $\tilde n(\Delta\log t)=\langle n(\log t,\log t+\Delta\log t)\,|\,\text{event at }t)\rangle$, where $n$ is the number of snaps at a given interval. Fig.\,\ref{fig:avalanches}a, left shows a clear distinction between the three regimes, corresponding to intra-cascade, inter-cascade (intra-avalanche) and inter-avalanche dynamics. The latter two exhibit clear power-law decays, while the transition between them marks the constant $\Theta$ defined above. Furthermore, the hierarchy is visible when computing $\tilde n(\Delta\log t)$ for cascades or for avalanches instead of single snaps.

These non-trivial correlations are also apparent in the distributions of log waiting times $\Delta\log t=\log t_{i}-\log t_{i-1}$ (Fig.\,\ref{fig:avalanches}b, left). Within avalanches, the distribution of waiting times between cascades exhibits a power law $P(\Delta\log t)\sim(\Delta\log t)^{-\alpha}$ with $\alpha\approx1$. 
Ref. \cite{korchinski2024microscopic} predicted this scaling for thermally activated, correlated barrier crossings.
Between avalanches, we find a second power law for the waiting time distribution $P(\Delta\log t)\sim(\Delta\log t)^{-\beta}$ with ${\beta\approx2}$. Together, the two regimes are reminiscent of the temporal correlations observed between subsequent earthquakes \cite{corral2003local,talbi2010mixed,li2024double}.


\section{Experimental validation}
Finally, we show that the insights gained from our analysis can be applied to creep experiments, in order to reveal complex correlations as predicted above. To this end, we revisit the results of Refs. \cite{shohat2023logarithmic,korchinski2024microscopic}, which considered the logarithmic compaction of crumpled thin sheets of Mylar under constant external load. This process was shown to be driven by thermal avalanches of discrete snap-trough instabilities, which could be individually identified using acoustic emission measurements \cite{shohat2023dissipation, lahini2023crackling,kramer1996universal, Houle1996}. 

We compute the conditional event rate in log-time for sequences of instabilities during logarithmic relaxation (Fig.\,\ref{fig:avalanches}a, right). Similarly to the model, we find a clear transition between two regimes characterized by different power laws (ultimately reaching a flat background noise level). Following this analogy, the transition clearly identifies a scale $\Theta^{\text{exp}}$ in log-time that allows separating intra- and inter-avalanche activity.

Using this scale, we define an avalanche $S_a^{\text{exp}}$ as a sequence of events with $\Delta\log t<\Theta^{\text{exp}}$ (see SI). To reduce noise, we ignore small avalanches of magnitude $S_a^{\text{exp}}\leq2$. The log waiting times statistics within and between avalanches are distributed as power laws with exponents $\approx-1$ and $\approx-2$ respectively (Fig.\,\ref{fig:avalanches}b, right). While the former scaling distribution was previously reported \cite{korchinski2024microscopic}, the latter is much more sensitive to the chosen value of $\Theta^{\text{exp}}$, thus demonstrating the strength of our analysis in revealing nontrivial correlations. {We note in passing that the $\sim\Delta^{-2}$ power law extends only a limited range of waiting times. Yet we stress that it is distinct from the intra-avalanche regime or from an exponential decay.}


\section{Discussion}
We studied the spatiotemporal structure of thermal avalanches during logarithmic creep. By examining correlations in space and time as well as the age-dependence of different behaviors, we showed how to systematically distinguish mechanical and thermal contributions. This revealed a hierarchical structure{, that is typically obscured in the raw data}: fast and compact mechanical bursts that facilitate each other's thermal activation, forming anomalously slow, system-spanning thermal avalanches. It is interesting to compare our results with the relaxation dynamics of glasses, so-called dynamical heterogeneities \cite{berthier2011dynamical}. Refs. \cite{candelier2009building,candelier2010spatiotemporal} observed a similar hierarchy of relaxation events that cluster both locally and globally. This hints to a potentially deeper connection between logarithmic creep in amorphous solids and slow relaxations in supercooled liquids and glass \cite{korchinski2024microscopic,tahaei2023scaling,scalliet2022thirty}.

The waiting time distributions we found (Fig.\,\ref{fig:avalanches}) show a striking similarity to the inter-event time distributions in a wide range of disordered systems, where two power-law regimes were observed \cite{corral2003local,talbi2010mixed,li2024double}. Thus, thermal avalanches suggest a potential mechanism for earthquake aftershock triggering \cite{marsan2024thermally}, as well as the intermittent dynamics during shear and compression of granular materials \cite{lherminier2019continuously,bares2017local}, porous materials \cite{baro2013statistical}, or wood \cite{makinen2015avalanches}. Interpreting the two power-law regimes as intra- and interavalanche correlations may shed light on their origin, as we demonstrated for the crumpled sheet experiments. 

{Finally, our bottom-up recipe for avalanche identification could be relevant for the analysis of other systems: from simulations of structural glasses, to earthquake data where clustering is notoriously difficult \cite{zaliapin2016global}. In both experiments and simulations of aging systems, mechanical cascades may not be accessible due to spatiotemporal resolution. Our analysis in nonetheless robust to this inaccessibility, as well as to noise (which is inherent to thermal avalanches), uneven coarse graining, or partial data. Particularly, the conditional rate analysis outperforms the abundantly used waiting time distribution analysis, and allows extracting characteristic correlation times from noisy measurements. 
}

\begin{acknowledgments}
We thank Daniel Korchinski, Marko Popovi\'{c}, and Matthieu Wyart for insightful comments.
This work was supported by the Israel Science Foundation grant 2117/22.
D.S. acknowledges support from the Clore Israel Foundation. V.R. acknowledges support from The Council of Higher Education and the Ministry of Immigration and Absorption. 
\end{acknowledgments}

\section*{Data Availability Statement}

Modified LAMMPS modules, and an example run script and output are available for download at \cite{lammps-download}.

\appendix
\section{Appendix A: Numerical simulations}

We simulated two-dimensional disordered networks consisting of $4\times10^4$ nodes of unit mass connected by $N~10^5$ bi-stable elastic bonds. Each bond has double-well potential $U_i(\ell_i)=a_4/4(\ell_i-\ell^{(0)}_i)^4 -a_2/2(\ell_i-\ell^{(0)}_i)^2$, where $\ell_i$ is the $i^{th}$ bond's length, $\ell_i^{(0)}$ is its rest length, and $a_2=2.5$, $a_4=1$ are network constants. The rest lengths were chosen randomly in the range $\ell^{(0)}\in[9,11]$, resulting in a random disordered network. We use over-coordinated networks with $\Delta Z=Z-Z_c\approx1.5$, where $Z$ is the mean number of neighbors per node, and $Z_c$ marks the isostatic point. More details on the network topology and its preparation can be found in Ref. \cite{shohat2022memory}.

Molecular dynamics was simulated in the canonical ensemble. An additional 12/6 Lennard-Jones potential with $\varepsilon_{LJ}=10$ and $\sigma_{LJ}=11.2246$ was introduced between pairs of non-neighboring nodes to avoid phantom behavior of the network. We use a Langevin thermostat at temperature $T=0.001$, and damping time $\tau_{damp}=4$.
All simulations were performed in LAMMPS \cite{thompson2022lammps} with modified \verb#bond_bpm#, \verb#bond_bpm_spring#, and \verb#bond_table# modules, which implement precise logging of bond snap events. Modified LAMMPS modules, and an example run script and output are available for download \cite{lammps-download}, see SI for details. 

To detect bond snapping, each bond potential was divided into three regions, or states: `short' ($\ell_i-\ell_i^{(0)}\leq-0.7$), `transition' ($-0.7<\ell_i-\ell_i^{(0)}\leq0.3$), and `long' ($\ell_i-\ell_i^{(0)}>0.3$).
The simulation tracks the state of each bond in the current time step, its state in the previous time step, and the last state different from the previous one (i.e. the state from which it came to the state at the previous timestep). We define a bond snap as a sequence of transitions from `long' to `transition' to `short' or \textit{vice versa}, but not `long' to `transition' to `long' or `short' to `transition' to `short'. In other words, a bond is considered to snap, when its current state is `long' (`short'), the previous state is `transition', and the state before that is `short' (`long'). Importantly, this instability tracking is done entirely on the fly in LAMMPS via the specific modules listed above, and does not require post-analysis of a stepwise trajectory. A graphical representation of the snap analysis is shown in the SI.

The initial network was prepared by the following equilibration procedure. First, the system was equilibrated without external stress at temperature $T=1$, which is high enough to overcome the bi-stable bond barrier and thus randomize states of all bonds. Second, the systems was gradually cooled to $T=0.001$ in $3\times10^4$ time units and then equilibrated at this temperature for $3\times10^4$ time units more to ensure the absence of bond snaps at low temperature.

To implement external stress conditions, a force $F=0.61$ acted on the boundaries of the network. Relaxation of the system under stress was simulated for $6\times10^6$ time units in 20 independent trajectories. In addition, 20 shorter simulations of $1.2\times10^6$ time units were performed for improved statistics results at shorter times. For every trajectory, the configuration of the network was written after each bond snap, allowing a detailed analysis of the deformation process.

\section{Appendix B: Effect of waiting time threshold $\Delta t^*$ on cascade properties\label{sec:appB}}

\begin{figure}[tbp]
 \centering
\includegraphics[width=0.49\linewidth, trim=10mm 0mm 26mm 0mm, clip]{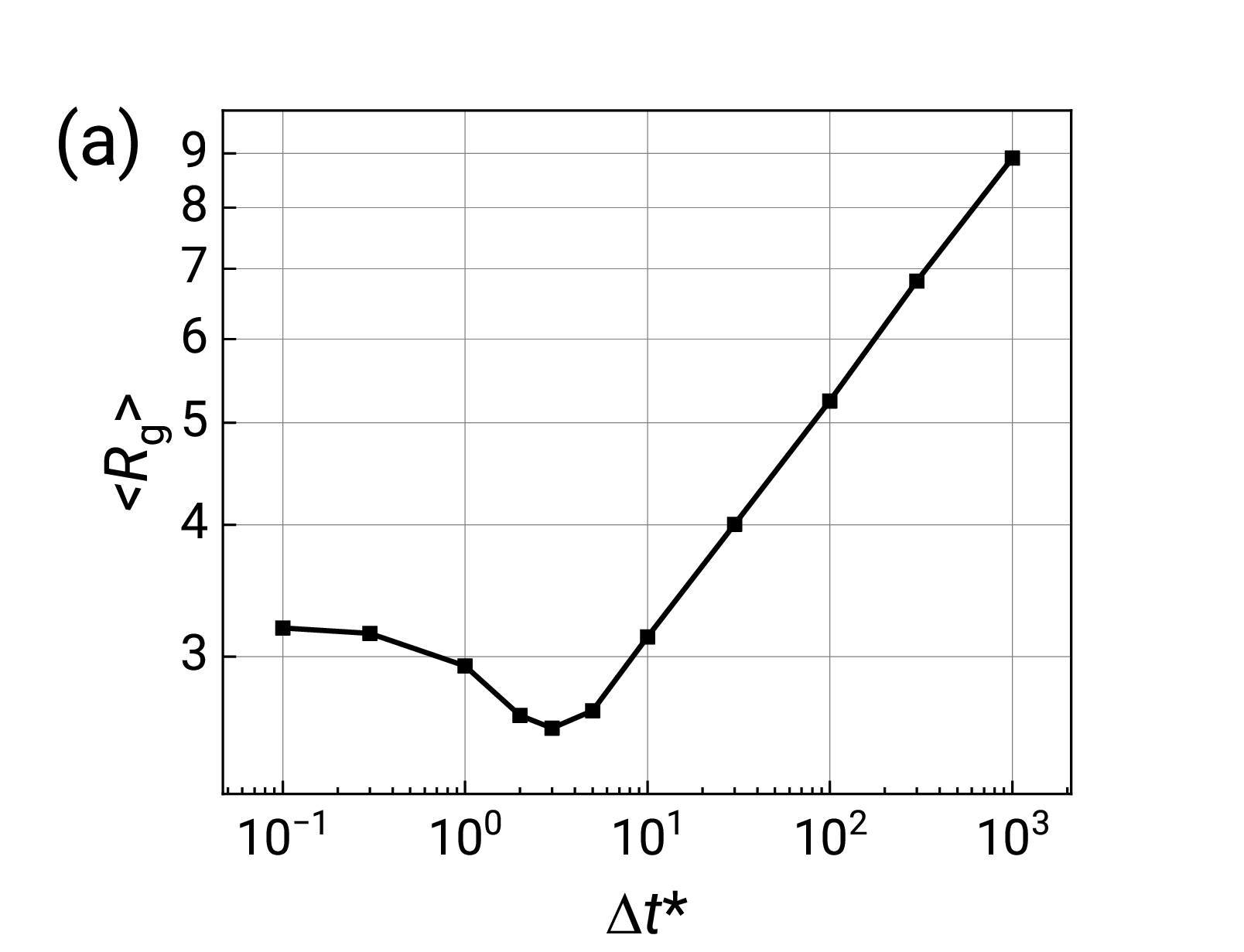}
\includegraphics[width=0.49\linewidth, trim=6mm 0mm 30mm 0mm, clip]{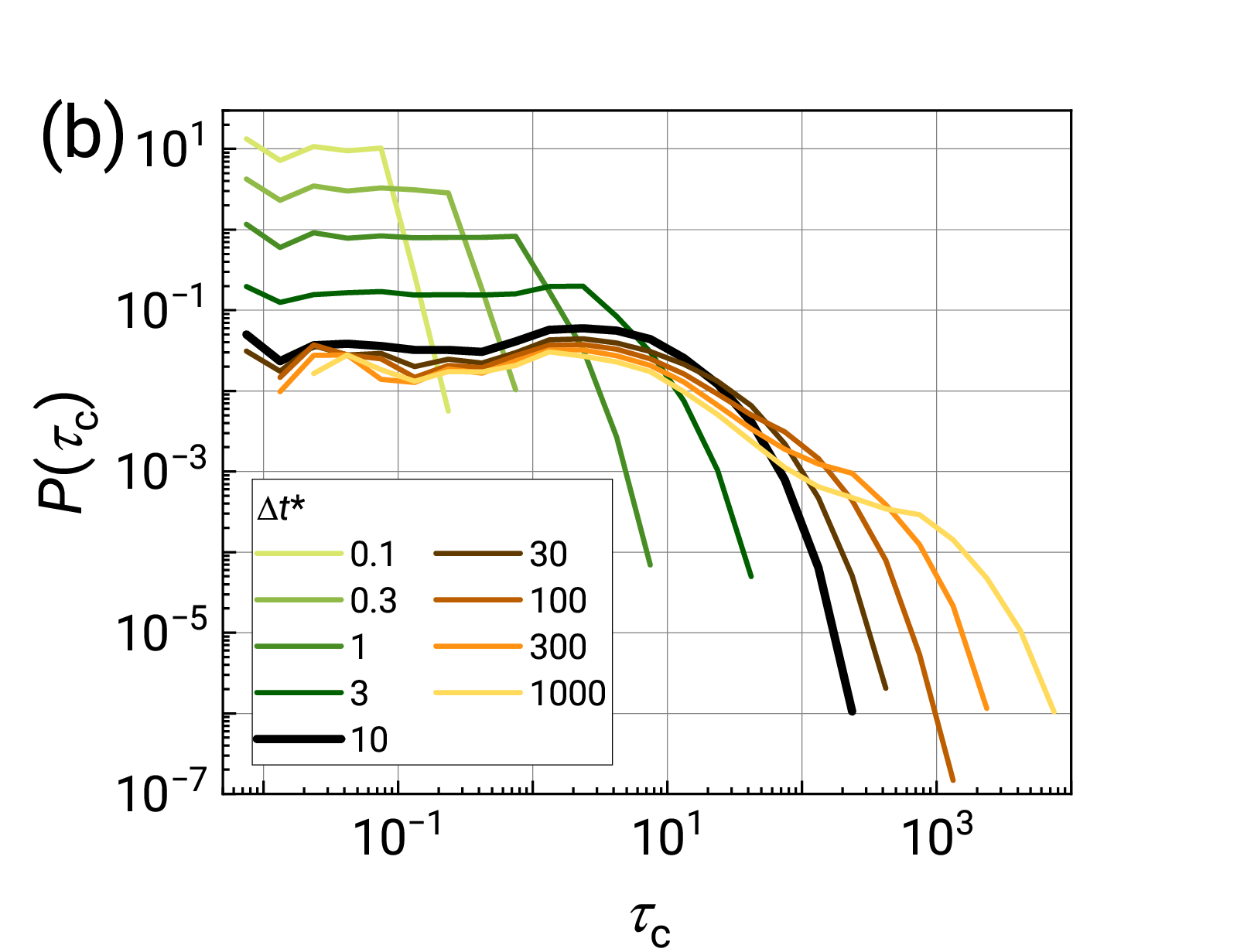}
\caption{\textbf{Mechanical cascades -- }
 (a) Average radius of gyration $\left<R_g\right>$ at various thresholds $\Delta t^*$. (b) Distribution of cascade durations $\tau_c$ at various thresholds $\Delta t^*$ for waiting times between instabilities.}
 \label{fig:cdurvardt}
\end{figure}

The waiting time threshold $\Delta t^*$ strongly affects the grouping of individual snaps into mechanical cascades. We analyzed the properties of the resulting cascades for various values of $\Delta t^*$ (see Fig.\,\ref{fig:cdurvardt}). 
At very small $\Delta t^*$,  the average radius of gyration $\left<R_g\right>$ is about 3 bond lengths, corresponding to an average distance to the next snap. With increasing $\Delta t^*$, first more local mechanical snaps get clustered into the same cascades, and $\left<R_g\right>$ decreases a bit due to better characterization of the compactness of mechanical cascades. However, from $\Delta t^*=3$ is raises up to the system size at large $\Delta t^*$, as more uncorrelated events from the whole system get clustered into one cascade. 

At the same time, too small values of $\Delta t^*$ result in cascades sharply interrupted too early, which gives a narrowed distribution of cascade duration $\tau_c$ with the sharp cutoff from the right. At the same time, too high values of $\Delta t^*$ lead to the occurrence of the secondary peak in the duration distribution. It directly denotes another type of process mixing with mechanical cascades at too chosen high values of $\Delta t^*$. 

In total, we believe the best choice of $\Delta t^*$ is between 3 and 10, as it allows the clustering algorithm to collect all the mechanically correlated snaps into single cascades, but doesn't let it to grab together events with different type of correlation.

%
%

%


\bibliography{bibli.bib}

\end{document}